\documentclass[twoside,twocolumn,english,pra,aps,superscriptaddress,showpacs]{revtex4-1}
\usepackage[T1]{fontenc}
\usepackage[latin9]{inputenc}
\usepackage{color}
\usepackage{babel}
\usepackage{textcomp}
\usepackage{amsmath}
\usepackage{amsthm}
\usepackage{graphicx}
\usepackage[unicode=true,pdfusetitle,
 bookmarks=true,bookmarksnumbered=false,bookmarksopen=false,
 breaklinks=false,pdfborder={0 0 0},backref=false,colorlinks=true]
 {hyperref}
\usepackage{breakurl}
\usepackage{epstopdf}
\usepackage{amssymb}
\makeatletter
\theoremstyle{plain}

\makeatother

\begin{document}

\preprint{This line only printed with preprint option}

\title{Landau-Zener-St\"{u}ckelberg Interferometry in $\mathcal{PT}$-symmetric Non-Hermitian models}

\author{Xin Shen}
\email{shenx@sustech.edu.cn}
\affiliation{Shenzhen Institute for Quantum Science and Engineering and Department of Physics, Southern University of Science and Technology, Shenzhen 518055, China }

\author{Fudong Wang}
\affiliation{Shenzhen Institute for Quantum Science and Engineering and Department of Physics, Southern University of Science and Technology, Shenzhen 518055, China }

\author{Zhi Li}
\email{lizhi_phys@m.scnu.edu.cn}
\affiliation{Guangdong Provincial Key Laboratory of Quantum Engineering and Quantum Materials, GPETR Center for Quantum Precision Measurement and SPTE, South China Normal University, Guangzhou 510006, China}

\author{Zhigang Wu}
\email{wuzg@sustech.edu.cn}
\affiliation{Shenzhen Institute for Quantum Science and Engineering and Department of Physics, Southern University of Science and Technology, Shenzhen 518055, China }
\affiliation{Center for Quantum Computing, Peng Cheng Laboratory, Shenzhen 518055, China}

\begin{abstract}
 We systematically investigate the non-Hermitian generalisations of the Landau-Zener (LZ) transition and the Landau-Zener-St\"{u}ckelberg (LZS) interferometry. The LZ transition probabilities, or band populations, are calculated for a generic non-Hermitian model and their asymptotic behaviour analysed. We then focus on non-Hermitian systems with a real adiabatic parameter and study the LZS interferometry formed out of two identical avoided level crossings. Four distinctive cases of interferometry are identified and the analytic formulae for the transition probabilities are calculated for each case. The differences and similarities between the non-Hermitian case and its Hermitian counterpart are emphasised. In particular, the geometrical phase originated from the sign change of the mass term at the two level crossings is still present in the non-Hermitian system, indicating its robustness against the non-Hermiticity. We further apply our non-Hermitian LZS theory to describing the Bloch oscillation in one-dimensional parity-time $(\mathcal{PT})$ reversal symmetric non-Hermitian Su-Schrieffer-Heeger model and propose an experimental scheme to simulate such dynamics using photonic waveguide arrays. The Landau-Zener transition, as well as the LZS interferometry, can be visualised through the beam intensity profile and the transition probabilitiess measured by the centre of mass of the profile.
\end{abstract}

\maketitle
\section{Introduction}\label{intro}
The transition between two energy levels of a particle driven across an avoided level crossing is a basic quantum dynamical process known as the Landau-Zener (LZ) transition \cite{landau1932,zener1932,stu,majorana1932}. Such a transition is often discussed under the time-dependent Hamiltonian
\begin{equation}
H(t)=Ft\sigma_z+m\sigma_x,
\label{LZhermitian}
\end{equation}
where $\sigma_{i}$ are the Pauli matrices, $F$ is the sweep velocity and $m$ is the mass or the gap parameter. The transition probability from the lower to upper band is given by the well-known LZ formula $P_{LZ} = e^{-2\pi \delta}$, where  $\delta =m^2/2F $ is  referred to as the adiabatic parameter. When a particle is driven through two such avoided level crossings successively, the transition probability from the lower to upper band depends not only on that of a single LZ transition $P_{LZ}$ but also on the total phase $\phi_t$ accumulated during the whole dynamical process \cite{lzs3,lzspiphase,lzs2,lzs4}, viz.,
\begin{align}
P_{-+} = 4P_{LZ}(1-P_{LZ})\sin^2\frac{\phi_t}{2}.
\label{LZS0}
\end{align}
As the path of the particle is split at the first level crossing and then merges at the second, such a physical scenario realises a type of interferometry in the energy-momentum space, namely the so-called Landau-Zener-St\"{u}ckelberg (LZS) interferometry. Due to their ubiquity and fundamental importance, the LZ transition and LZS interferometry have been studied experimentally in a wide range of physical systems such as the nano-structure \cite{lznano1,lznano2,lznano3}, the Bose-Einstein condensates \cite{lz1,lzbec2,lzbec4} and the superconducting qubits \cite{lzqubit1,lzqubit2}, to name just a few.

 Conceptually, efforts have  also been put forward to extend the simple paradigm of the two-level LZ problem to more complex scenarios. One notable direction is the proposal of the multi-state LZ transition involving more energy levels than two, where examples include the so-called bow-tie model \cite{mlz2,mlz3}, the Demkov-Osherov \cite{mlz1} model and other models in realistic systems \cite{lzcqed,lztc}. Another direction, which is the focus of this work, is to consider the LZ transitions in non-Hermitian systems~\cite{vitanov1997,lzdecay,lznHer,Reyes_2012,nonlzR}. Non-Hermitian systems, particularly those with parity and time reversal ($\mathcal{PT}$) symmetry \cite{ptnHer,ptnphy}, have attracted growing attention as a result of the development of topological band theory \cite{tbtnHer,topoclanHer,wangzhong1,wangzhong2,bbcnHer,hallconduc1,hallconduc2,hallconduc3}.  For these systems, the non-Hermitian Hamiltonian generally possesses real energy spectrum except when the $\mathcal{PT}$ symmetry is spontaneously broken \cite{ptssb1}. One way to realise such non-Hermitian models experimentally is to use optical setups and introduce optical gain and loss  \cite{gainandloss,Weimann2016}.  For instance, a $\mathcal{PT}$ symmetric non-Hermitian Su-Schrieffer-Heeger (SSH) model has been implemented with photonic waveguide arrays \cite{Weimann2016}, in which the existence of topological interface states is demonstrated.

Motivated by the interests in both the LZ problem and the non-Hermitian physics, we investigate in this paper the LZ transition and the LZS interferometry in non-Hermitian systems~\cite{hatanoNelson,longhi2015a,imaginary_gauge1,imaginary_gauge2}. We first provide a complete solution of the LZ transition  for a generic two-band non-Hermitian model, which can be viewed as generalisations of earlier results found in Ref.~\cite{vitanov1997,lzdecay,lznHer}. The central result of our work is an extension of the transition probability given in Eq.~(\ref{LZS0}) for Hermitian systems to that for non-Hermitian systems with a real adiabatic parameter. As a concrete application of our analytic results on the non-Hermitian LZS interferometry, we further examine the  $\mathcal{PT}$-symmetric non-Hermitian SSH model and demonstrate that our results provide an accurate description of the exact LZS dynamics, even in the $\mathcal{PT}$-symmetry-breaking regime. Finally, we propose an experimental scheme to test our theory using the wave packets in photonic waveguide arrays. This scheme is made possible due to the similarity between the Schr\"{o}dinger's equation and the scalar paraxial wave equation for optical waves, which allows for a classical simulation of quantum effects such as the \emph{Zitterbewegung} \cite{opticalzb}, Klein tunneling \cite{opticalkt} and Bloch oscillation \cite{bo1,bo2}. Both the LZ transition and LZS interferometry are shown to be experimentally observable through the measurements of the beam intensity profile.

The rest of the paper is organised as follows: In Sec.~\ref{s2}, we solve the general non-Hermitian LZ model and analyse the transition probabilities and their asymptotic behaviours.  We then adopt the adiabatic-impulse model to analyse the LZS interferometry in Sec.~\ref{s3} and derive analytic expressions for the final lower and upper band populations. These analytic results are compared to the exact numerical simulation of the dynamics in the context of a non-Hermitian SSH model. Finally, an experimental proposal to simulate the relevant dynamics in such a non-Hermitian SSH model is discussed in Sec.~\ref{s4} and some concluding remarks are given in Sec.~\ref{s5}.

\section{non-Hermitian Landau-Zener transition}
\label{s2}
\subsection{General solutions}
Building on the Hermitian LZ Hamiltonian given in Eq.~(\ref{LZhermitian}), we consider the following generic non-Hermitian LZ Hamiltonian
\begin{equation}
\label{glz}
H(t)=(Ft+i\kappa)\sigma_z+(m+im')\sigma_x+(n+in')\sigma_y,
\end{equation}
where the sweep velocity $F$ is taken to be positive and  $\kappa$, $(m,n)$, $(m',n')$ are all real parameters.

To facilitate the discussion of the non-Hermitian LZ transition, we first clarify the band notations. The left and right adiabatic eigenstates are defined as $|u^{L/R}_\pm(t)\rangle$ satisfying
\begin{equation}
\begin{split}
H(t)|u^R_\pm(t)\rangle&= E_{\pm}(t)|u^R_\pm(t)\rangle\\
\langle u_{\pm}^L(t)|H(t)&=  E_{\pm}(t)\langle u^L_\pm(t)|,
\end{split}
\end{equation}
where $E_{+}(t)$ and $E_{-}(t)$, two roots of the equation $\text{det}(H(t) -E I) = 0$, are the upper and lower band adiabatic dispersion respectively.  We note that $E_{\pm}$ can in principle be complex for a non-Hermitian system. It can be shown that $\langle u^L_{\pm}|u^R_{\mp}\rangle=0$ as long as $E_+\neq E_-$. Assuming no degeneracies, i.e., $\langle u^L_{\pm}|u^R_{\pm}\rangle\neq0$, we can construct the projection operator
\begin{equation}
{\mathcal P}_{\pm}=\frac{1}{\langle u^L_{\pm}|u^R_{\pm}\rangle} |u^R_{\pm}\rangle\langle u^L_{\pm}|
\end{equation}
such that the Hamiltonian can be decomposed as $H=\sum_{\pm}E_{\pm}{\mathcal P}_{\pm}$.
Supposing the initial state is prepared in the ground state $|u^R_-(-T)\rangle$, then the time-evolved state $|\psi_-(t)\rangle\equiv U(t,-T)|u^R_-(-T)\rangle$ is derived by solving the Schr\"{o}dinger's equation. The non-Hermitian generalisation of the transition probablities are given by
\begin{equation}
\label{bandpop}
\begin{split}
P_{-+}(t)&=\frac{1}{|\langle u^L_{+}|u^R_{+}\rangle|^2}|\langle u^L_+(t)|\psi_-(t)\rangle|^2,\\
P_{--}(t)&=\frac{1}{|\langle u^L_{-}|u^R_{-}\rangle|^2}|\langle u^L_-(t)|\psi_-(t)\rangle|^2.
\end{split}
\end{equation}
 If the initial state is prepared in $|u^R_+(-T)\rangle$ which evolves into $|\psi_+(t)\rangle\equiv U(t,-T)|u^R_+(-T)\rangle$, we can similarly define
\begin{equation}
\label{bandpop1}
\begin{split}
P_{+-}(t)&=\frac{1}{|\langle u^L_{-}|u^R_{-}\rangle|^2}|\langle u^L_-(t)|\psi_+(t)\rangle|^2,\\
P_{++}(t)&=\frac{1}{|\langle u^L_{+}|u^R_{+}\rangle|^2}|\langle u^L_+(t)|\psi_+(t)\rangle|^2.
\end{split}
\end{equation}
Adopting the definition in the Hermitian case, the LZ transition probability is given by
\begin{align}
 P_{LZ}\equiv \lim_{T\rightarrow \infty}P_{-+}(T).
 \end{align}
Here the term transition probability is used loosely, since $P_{-+}$ and $P_{--}$ in principle can exceed unity due to the fact that the evolution is no longer unitary. For the same reason, we also have $P_{--}(t) \neq 1- P_{-+}(t) $ for non-Hermitian Hamiltonians. Later we will also use the term band population to refer to these quantities.

\begin{figure}[t]
\begin{center}
\includegraphics[width=0.4\textwidth]{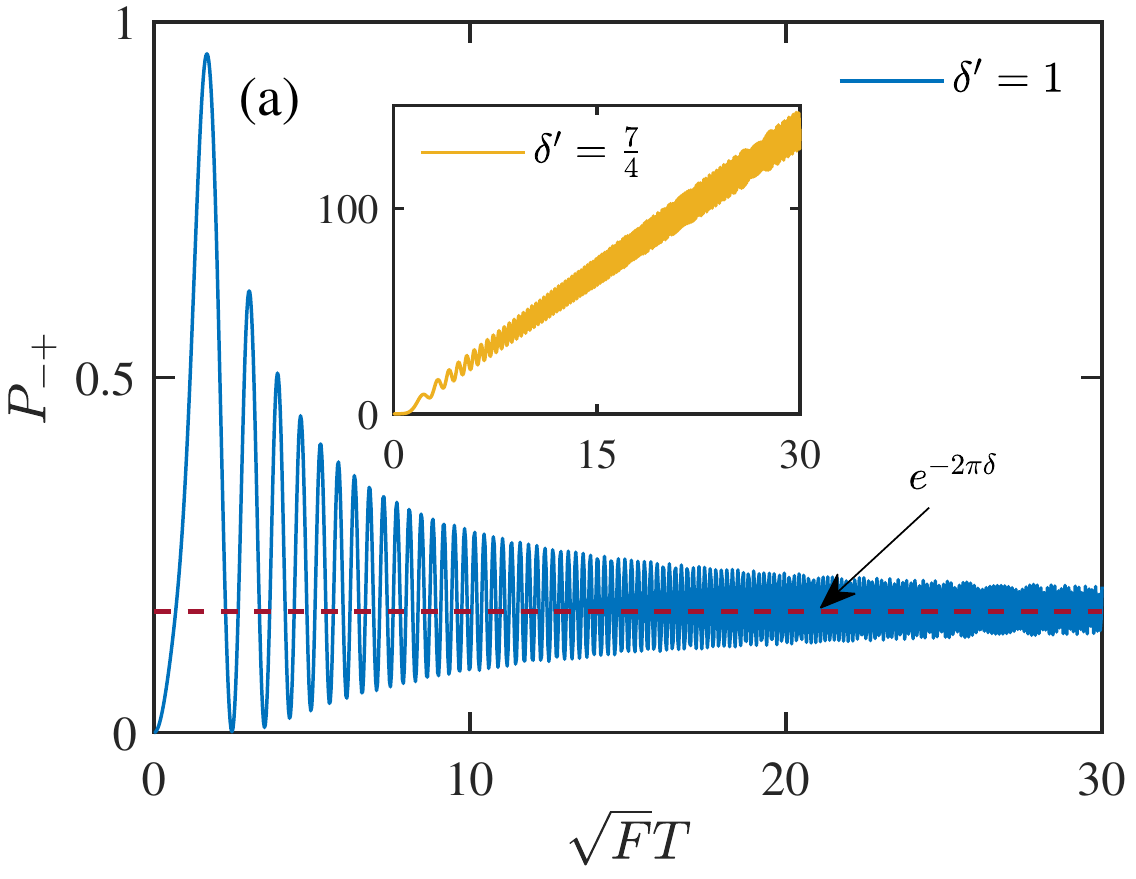}\\
\includegraphics[width=0.4\textwidth]{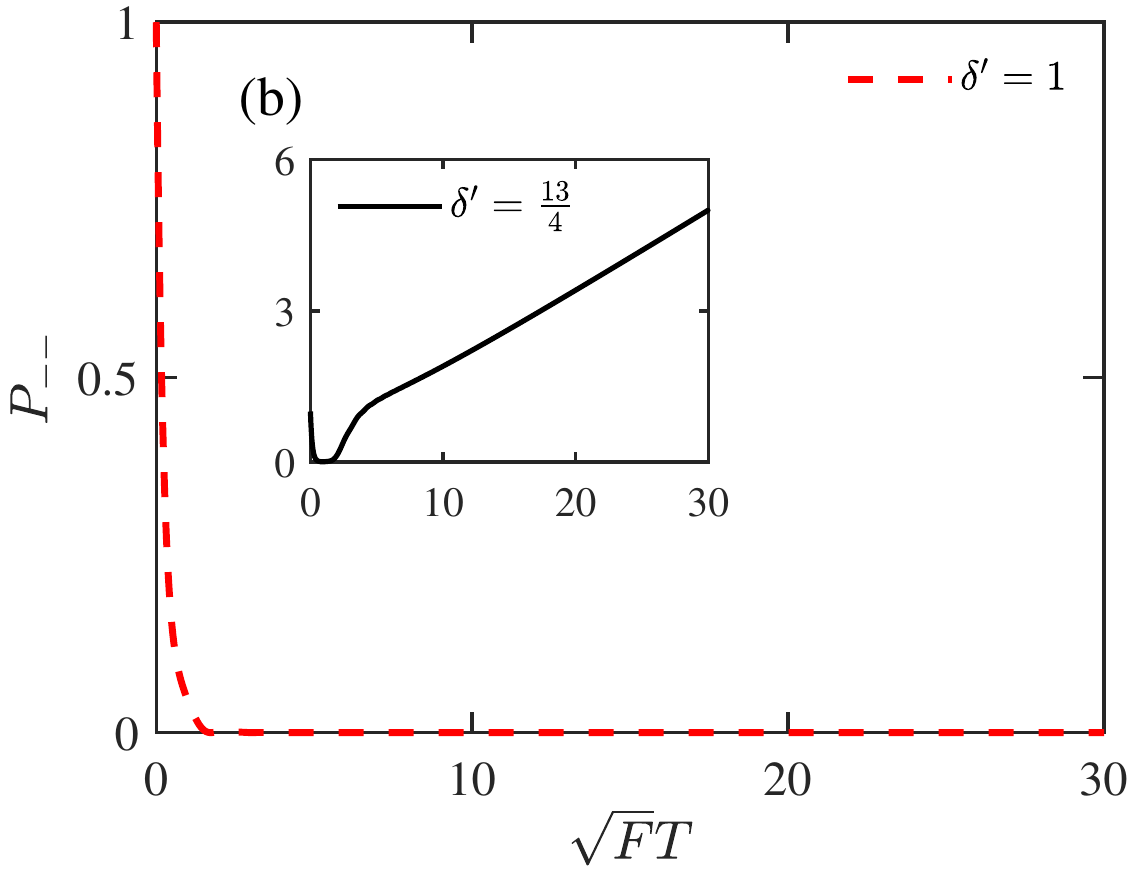}
\caption{(Color online) Numerical calculations of the transition probability (a) $P_{-+}$ and (b) $P_{--}$ defined in Eq.~\eqref{bandpop} for various $\delta'$ parameters. The numerical results agree with the analytical predictions in Eqs.~\eqref{pmpasym} and \eqref{pmmasym} in the large $T$ limit. In both (a) and (b), we take $\delta=9/32$ and $m=n'$. }
\label{default}
\end{center}
\end{figure}
We now solve the Schr\"{o}dinger equation governed by the Hamiltonian Eq.~(\ref{glz}). This is most conveniently done in the diabatic basis, namely the eigenstates of $\sigma_z$. For the two-level system, we also refer to the state  $[1,0]^T$ as the spin-up state and $[0,1]^T$  as spin-down state.  The $\kappa$ parameter can be absorbed into the time argument by defining a complex time and the solution for a finite $\kappa$ can be derived by a subsequent  analytical continuation of the solution with $\kappa=0$ \cite{lzdecay}. The problem is thus reduced to finding the solution for the case of $\kappa=0$ and the corresponding Schr\"{o}dinger equation reads
\begin{align}
i\partial_ta&=Ft a+\left[m +im'-i(n+in')\right]b\\
i\partial_tb&=\left[m +im'+i(n+in')\right]a-Ftb,
\end{align}
where the wavefunction is written in the form of $\psi(t)=[a(t), b(t)]^T$.
As the usual Landau-Zener problem, this can be written in the form of the second order Weber equations
\begin{align}
\label{weber1}
\ddot{a}+\left[i+2(\delta+i\delta ')+z^2\right]a&=0\\
\ddot{b}+\left[-i+2(\delta+i\delta ')+z^2\right]b&=0,
\label{weber2}
\end{align}
where the dot denote the derivative with respect to $z\equiv\sqrt{F}t$. In contrast to the Hermitian case, the adiabatic parameter here is a complex number and we define the real and imaginary adiabatic  parameters as  $\delta=(m ^2-m'^2+n ^2-n'^2)/2F$ and $\delta '=(m m'+n n')/F$, respectively. The solution to Eqs.~(\ref{weber1})-(\ref{weber2}) is a linear combination of the parabolic cylinder functions $D_p$
\begin{align}
\label{solution}
a&=\sum_{\pm}A_{\pm}D_{p}(\pm\sqrt{2}e^{i\pi/4}z)\\
b&=\sum_{\pm}B_{\pm}D_{p-1}(\pm\sqrt{2}e^{i\pi/4}z),
\end{align}
where $A_\pm$ and $B_\pm$ are the coefficients depending on the initial state and $p\equiv-i\delta+\delta '$. Inserting the solution to the first order differential equations, we find the relations
\begin{equation}
\label{abcoeff}
\begin{split}
A_\pm&=\pm C_{AB}B_\pm,\\
C_{AB}&=\frac{(m +im')-i(n +in')}{\sqrt{2F}(\delta+i\delta ')}e^{-i\frac{\pi}{4}}.
\end{split}
\end{equation}
The evolution matrix connecting the initial and final state is determined by
\begin{equation}\label{evolu1}
\begin{bmatrix}
a(t_f) \\
b(t_f)
\end{bmatrix}=U(t_f, t_i)
\begin{bmatrix}
a(t_i) \\
b(t_i)
\end{bmatrix}.
\end{equation}
Combining Eqs. \eqref{solution} -\eqref{abcoeff} we obtain the elements of the evolution matrix as
\begin{equation}
\label{ut}
\begin{split}
U_{11}&=\frac{D_{p}(\tau_f)D_{p-1}(-\tau_i)+D_{p}(-\tau_f)D_{p-1}(\tau_i) }{D_{p}(\tau_i)D_{p-1}(-\tau_i)+D_{p}(-\tau_i)D_{p-1}(\tau_i)}\\
U_{12}&=\frac{D_{p}(\tau_f)D_{p}(-\tau_i)-D_{p}(-\tau_f)D_{p}(\tau_i) } {D_{p}(-\tau_i)D_{p-1}(\tau_i)+D_{p}(\tau_i)D_{p-1}(-\tau_i)}C_{AB}\\
U_{21}&=\frac{D_{p-1}(\tau_f)D_{p-1}(-\tau_i)-D_{p-1}(-\tau_f)D_{p-1}(\tau_i) }{D_{p}(\tau_i)D_{p-1}(-\tau_i)+D_{p}(-\tau_i)D_{p-1}(\tau_i)}\frac{1}{C_{AB}}\\
U_{22}&=\frac{D_{p}(-\tau_i)D_{p-1}(\tau_f)+D_{p}(\tau_i)D_{p-1}(-\tau_f) }{D_{p}(-\tau_i)D_{p-1}(\tau_i)+D_{p}(\tau_i)D_{p-1}(-\tau_i)},
\end{split}
\end{equation}
where $\tau_{i,f}=\exp(i\pi/4)\sqrt{2F}t_{i,f}$.  For the previously defined band populations in Eq.~(\ref{bandpop}), we are interested in the large $T$ behaviour of the band populations $P_{-+}(T)$ and $P_{--}(T)$, which can be obtained by making use of the following asymptotical expressions of the parabolic cylinder functions
\begin{equation}
\label{at}
\begin{split}
&D_p\left (\sqrt{2}e^{i\pi/4}z_a\right )\sim e^{-i\left[\Phi(z_a)-\frac{\pi}{4}\delta '\right]+\delta '\ln\sqrt{2}z_a+\frac{\pi}{4}\delta}\\
&\times\left[1+\frac{ip(p-1)}{4z_a^2}-\frac{p(p-1)(p-2)(p-3)}{32z_a^4}\right],\\
&D_p\left (-\sqrt{2}e^{i\pi/4}z_a\right )\sim e^{-i\left[\Phi(z_a)+\frac{3\pi}{4}\delta '\right]+\delta '\ln\sqrt{2}z_a-\frac{3\pi}{4}\delta}\\
&\times\left[1+\frac{ip(p-1)}{4z_a^2}-\frac{p(p-1)(p-2)(p-3)}{32z_a^4}\right]+\\
&\frac{\sqrt{2\pi}}{\Gamma(i\delta-\delta ')}e^{i\left[\Phi(z_a)-\frac{\pi}{4}(\delta '+1)\right]-(\delta '+1)\ln\sqrt{2}z_a-\frac{\pi\delta}{4}}\\
&\times\left[1-\frac{ip(p-1)}{4z_a^2}-\frac{p(p-1)(p-2)(p-3)}{32z_a^4}\right],
\end{split}
\end{equation}
where $z_a\equiv\sqrt{F}T\rightarrow +\infty$, $\Phi(z_a) = z_a^2/2+\delta\ln\sqrt{2}z_a$ and $\Gamma$ denotes the Gamma function. Using these expressions in Eq.~(\ref{ut}) and Eq.~(\ref{bandpop}) we find
\begin{align}\label{pmpasym}
P_{-+}(T) \sim \left \{\begin{array}{cc} e^{-2\pi\delta} & \qquad  |\delta'| < 3/2 \\ T^{4|\delta'|-6}  & \qquad |\delta'| > 3/2  \\  f_0(\delta) + f(T) & \qquad    |\delta'| = 3/2\end{array}\right.
\end{align}
where $f_0(\delta)=\frac{1+e^{-2\pi\delta}}{\delta^2+1/4} + e^{-2\pi\delta}$ and $f(T)=-2e^{-\pi\delta}\sqrt{\frac{1+e^{-2\pi\delta}}{\delta^2+1/4}}\sin[2\Phi-\arg\Gamma(i\delta-{1}/{2}]$ is an oscillatory function of $T$. Similarly we find
\begin{align}\label{pmmasym}
P_{--}(T) \sim \left \{\begin{array}{cc} 0 & \qquad  |\delta'-3/2| < 3/2 \\ T^{4|\delta'-3/2|-6}  & \qquad |\delta'-3/2| > 3/2  \\ g_{0,3}(1-e^{-2\pi\delta}) &   \qquad  |\delta'-3/2| = 3/2\end{array} \right.
\end{align}
where $g_0=\frac{m^2+n^2-m^{'2}-n^{'2}}{(m+n')^2+(m'-n)^2}$ and $g_3 =\frac{\delta^2+9}{\delta^2(\delta^2+1)(\delta^2+4)}g_0$ correspond to $\delta'=0$ and $\delta'=3$, respectively. We see that the large $T$ behaviour of the band populations depends crucially on the parameter $\delta'$, i.e., the imaginary part of the adiabatic parameter. In particular, in order for both band populations to be finite, $\delta'$ shall lie within the interval  $(0,3/2)$ where $P_{-+}(T)$ attains the same asymptotic value as the Hermitian case. In Fig.~\ref{default}, we show the numerically calculated $P_{-+}(t)$ and $P_{--}(t)$ for various values of $\delta'$, which exhibit the asymptotic behaviour predicted by our analytic expressions in Eqs.~(\ref{pmpasym})-({\ref{pmmasym}}). 

Lastly, we discuss the choice of the basis in defining the transition probabilities for non-Hermitian systems. Previously we have chosen the adiabatic basis for the definitions in Eqs.~(\ref{bandpop})-(\ref{bandpop1}). Since the adiabatic eigenstates approach the diabatic ones in the  $T\rightarrow \infty$ limit, one may be led to think that the transition probabilities would be independent of the choice of the basis, just as the Hermitian case. However, this is not the case for non-Hermitian systems. This is because the non-unitary evolution matrix elements may become divergent in the $T\rightarrow \infty$ limit, in which case the transition probability in the adiabatic basis depends crucially on the asymptotic behaviour of adiabatic eigenstates. This is the reason why 
 the transition probabilities defined in terms of the diabatic basis are not necessarily the same as those defined in terms of the adiabatic basis. To see this more explicitly, we show below the asymptotic behaviuor of the  transition probability defined in diabatic basis, i.e., 
$P^d_{-+}(T)=\left|U_{11}(T,-T)\right|^2$ and 
$P^d_{--}(T)=\left|U_{21}(T,-T)\right|^2$,
\begin{align}\label{pmpasymd}
P^d_{-+}(T) \sim \left \{\begin{array}{cc} e^{-2\pi\delta} & \qquad  |\delta'| < 1/2 \\ T^{4|\delta'|-2}  & \qquad |\delta'| > 1/2  \\ 1 + 2e^{-2\pi\delta} + f^d(T) & \qquad    |\delta'| = 1/2\end{array}\right.
\end{align}
and 
\begin{align}\label{pmmasymd}
P^d_{--}(T) \sim \left \{\begin{array}{cc} 0 & \qquad  |\delta'-1/2| < 1/2 \\ T^{4|\delta'-1/2|-2}  & \qquad |\delta'-1/2| > 1/2  \\ g_{0,1}(1-e^{-2\pi\delta}) &   \qquad  |\delta'-1/2| = 1/2\end{array} \right.
\end{align}
where $f^d(T)=-2e^{-\pi\delta}\sqrt{1+e^{-2\pi\delta}}\sin[2\Phi-\arg\Gamma({1}/{2}+i\delta)]$, $g_0$ was defined earlier and  $g_1=\frac{(m'+n)^2+(m-n')^2}{m^2+n^2-m^{'2}-n^{'2}}$ corresponding the case $\delta'=1$. Comparing Eqs.~(\ref{pmpasymd})-(\ref{pmmasymd}) to Eqs.~(\ref{pmpasym})-(\ref{pmmasym}), we find that only when $\delta'=0$, in which case the evolution matrix is convergent, the asymptotic transition probabilities defined in the two bases are identical. The rest of the paper will be devoted to studying this type of non-Hermitian systems.

\subsection{Systems with real adiabatic parameters}
In the case of $\delta' = 0$, the complex coefficients in front of the $\sigma_x$ and $\sigma_y$ terms in the Hamiltonian Eq.~(\ref{glz}) have a relative $\pi/2$ phase. Choosing a gauge under which the gap parameter is real, we arrive at a Hamiltonian 
\begin{equation}
\label{lzham}
h(t)=Ft\sigma_z+m\sigma_x+i\gamma\sigma_y,
\end{equation}
where we have replaced the previously used notation $n'$ by $\gamma$, following the commonly used notation in the literature. The Hamiltonian has the  $\mathcal{PT}$ symmetry, i.e., $\mathcal{PT}~h(t)~(\mathcal{PT})^{-1}=h(-t)$, where $\mathcal{PT}=\sigma_x\mathcal{K}$ and $\mathcal{K}$ is the complex conjugate operator.  The adiabatic spectrum of the above Hamiltonian is
\begin{align}
E_{\pm} (t) = \pm \sqrt{F^2t^2 + m^2 - \gamma^2}
\label{Epm}
\end{align}
and the real adiabatic parameter is given by
\begin{align}
\delta  = (m^2- \gamma^2)/2F.
\end{align}
The energy spectrum keep real except when  $\gamma>m$, which indicates the spontaneous broken of $\mathcal{PT}$ symmetry \cite{ptnHer,ptssb1}. From Eq. \eqref{ut} we obtain the following asymptotic form of the evolution matrix elements
\begin{equation}
\label{plz4}
\begin{split}
U_{11}=&U_{22}=e^{-\pi\delta}\\
U_{12}=& {\rm sgn}(m+\gamma)\sqrt{\frac{m+\gamma}{m-\gamma}\left (1-e^{-2\pi\delta}\right )}e^{-i\phi_s}\\
U_{21}=&{\rm sgn}(-m+\gamma)\sqrt{\frac{m-\gamma}{m+\gamma}\left (1-e^{-2\pi\delta}\right )}e^{i\phi_s}\\
\end{split}
\end{equation}
where $\phi_s\equiv \pi/4+\delta(\ln|\delta|-1)-\arg\Gamma(1+i\delta)$ is the Stokes phase. Note that the dynamical phase accumulated from the adiabatic evolution away from the crossing is neglected for the moment. The above results are consistent with those given in Ref. \cite{ptlz,lznHer}.
The band population can be readily obtained as
\begin{equation}
\label{nlz}
\begin{split}
P_{-+}(T)&= P_{+-}(T)=P_{LZ}\\
P_{--}(T)&=\frac{m-\gamma}{m+\gamma}(1-P_{LZ})\\
P_{++}(T)&=\frac{m+\gamma}{m-\gamma}(1-P_{LZ}),
\end{split}
\end{equation}
 in the $T \rightarrow  \infty$ limit, where $P_{LZ} = e^{-2\pi\delta}$.

 Unsurprisingly, the transition probabilities are no longer solely dependent on the adiabatic parameter $\delta$.  Although the expression for $P_{LZ}$ is the same as that for the corresponding Hermitian counterpart, the results of the non-Hermitian LZ transition have several marked differences. First, the addition of the non-Hermitian term reduces the adiabatic parameter $\delta=(m^2-\gamma^2)/2F$ and thus enhances the transition probability from one band to the other. In fact, the adiabatic parameter can take negative values as $\gamma$ increases such that the transition probability is greater than unity. Furthermore, we see that $P_{--} \neq P_{++}$, which means that the probability to stay in the same band after the LZ transition depends on the initial state. This asymmetrical effect is most pronounced at $\gamma=\pm m$ for which the adiabatic parameter vanishes. In the Hermitian system, the vanishing adiabatic parameter leads to a perfect transmission, meaning that the particle remains in the same spin state after the transition; this is true regardless of whether the initial state is spin-up or spin-down, i.e., the perfect transmission is symmetrical with respect to the initial state. In the non-Hermitian system, however, it is not the case. Take $\gamma = m$ for example. The transition matrix reads
  \begin{equation}
U=
\begin{pmatrix}
1 & 2m\sqrt{{\pi}/{F}}e^{-i\phi_s} \\
0 &1
\end{pmatrix},
\end{equation}
which shows that perfect transmission occurs for the spin-up state but not for the spin-down state. The situation is reversed for $\gamma = -m$.

Finally, we end this section by discussing the time scale of the non-Hermitian LZ transition.  Following Ref.~\cite{jumptime} we define the LZ transition time as
\begin{equation}
\tau_{{LZ}}=\lim_{T\rightarrow \infty}\frac{P_{-+}(T)}{P'_{-+}(0)},
\end{equation}
which is roughly the time taken for the interband transition probability to reach its asymptotic value from zero. Using the previously obtained transition matrix Eq.~(\ref{ut}), we find
\begin{equation}
\tau_{{LZ}}=\frac{|1-P_{LZ}|}{\sqrt{2\delta(1-P_{LZ})}\cos{\chi(\delta)}},
\end{equation}
where $\chi(\delta)=\pi/4-\arg\Gamma(\frac{1}{2}+i\frac{\delta}{2})-\arg\Gamma(1-i\frac{\delta}{2})$. The transition times in various limiting situations are summarised as
\begin{equation}
\label{taulz}
\tau_{LZ}=
\begin{cases}
\frac{1}{\sqrt{-2\delta}}e^{-\pi\delta}  &\delta\ll-1\\
\sqrt{2\pi}  &|\delta|\ll1 \\
2\sqrt{2\delta} &\delta\gg1.
\end{cases}
\end{equation}
We note that for $\delta > 0 $ the non-Hermitian LZ transition time shares a similar form as that for the Hermitian counterpart. The case of $\delta<0$, however, has no equivalence in the Hermitian case. In fact, the LZ transition time grows exponentially as the adiabatic parameter decreases, reflecting the unique nature of the presence of the spontaneous $\mathcal{PT}$ symmetry breaking and the existence of exceptional points.

\section{Non-Hermitian Landau-Zener-St\"{u}ckelberg interferometry}
\label{s3}
In this section, we examine the quantum dynamics governed by a time-dependent non-Hermitian Hamiltonian whose adiabatic energy spectrum $E_{\pm} (t)$ contains two identical avoided level crossings (see Fig.~\ref{lzs} for example). For the purpose of generality, we do not need to specify the precise form of this Hamiltonian except to assume that it can be be well approximated by Eq.~(\ref{lzham}) in the vicinity of the avoided crossings. The validity of this assumption will be examined later for the specific Hamiltonian of interest. As we mentioned earlier, the splitting and merging of the paths traversed by the particle realises the Landau-Zener-St\"{u}ckelberg interferometry.  We shall focus on a regime where the LZS dynamics can be viewed as consisting of two parts, namely sudden LZ transitions at the avoided crossings and adiabatic evolutions away from them. Such a treatment of the dynamics is referred to as the adiabatic-impulse theory and is adopted in Sec.~\ref{sa} to determine the band populations for the non-Hermitian LZS interferometry. In Sec.~\ref{sb}, a concrete example of this kind of dynamics is studied in the non-Hermitian SSH model, where the electron at the bottom of the lower band is driven by a static electric field and undergoes the Bloch oscillation. We show that the analytic results from the adiabatic-impulse theory agrees rather well with the exact dynamics determined from the numerics.
\subsection{Adiabatic-impulse theory}
\label{sa}
Let's consider two identical avoided level crossings of the type in Eq.~(\ref{Epm}) at $t=t_1$ and $t= t_2>t_1$. We  take the local Hamiltonian at $t_1$ to be
\begin{equation}
\begin{split}
\label{effham1}
h_1(t)=F(t-t_1)\sigma_z+m\sigma_x+i\gamma \sigma_y,
\end{split}
\end{equation}
where $F$ is assumed to be positive without loss of generality. Four possible local Hamiltonians at $t_2$ are allowed by the energy spectrum in Eq.~(\ref{Epm}), namely
\begin{equation}
\begin{split}
\label{effham2}
h_2(t)=-F(t-t_2)\sigma_z \pm  m\sigma_x \pm i\gamma \sigma_y,
\end{split}
\end{equation}
where the negative sign in front of the $\sigma_z$ term ensures that the adiabatic eigenstates to the right of the first crossing are connected smoothly to those to the left of the second. Later we shall classify different types of the LZS interferometry in terms of the combinations of the coefficients $(\pm m,\pm \gamma)$ of the $\sigma_x$ and $\sigma_y$ terms in $h_2(t)$.

The LZS interferometry is most conveniently studied in the adiabatic basis. Away from $t=t_{1,2}\pm T$ the non-Hermitian term can be neglected and the left and right adiabatic eigenstates will not be distinguished. Thus we arrive at the following adiabatic basis states for the Hamiltonian $h_1(t)$
\begin{equation}
\begin{split}
 |u_{+}(t_1-T)\rangle&=
 \begin{pmatrix}
 0\\
 1
 \end{pmatrix},
 |u_{+}(t_1+T)\rangle=
\begin{pmatrix}
1 \\ 0
\end{pmatrix}\\
|u_{-}(t_1-T)\rangle&=
\begin{pmatrix}
1 \\ 0
\end{pmatrix},
|u_{-}(t_1+T)\rangle=
 \begin{pmatrix}
 0\\
 1
 \end{pmatrix}
\end{split}
\end{equation}
where, as usual, $+$ and $-$ denote the upper and lower bands respectively.
Similarly for the Hamiltonian $h_2(t)$, we find
\begin{equation}
\begin{split}
 |u_{+}(t_2-T)\rangle&=
 \begin{pmatrix}
 1\\
 0
 \end{pmatrix},
 |u_{+}(t_2+T)\rangle=
\begin{pmatrix}
0 \\ 1
\end{pmatrix}\\
|u_{-}(t_2-T)\rangle&=
\begin{pmatrix}
0 \\ 1
\end{pmatrix},
|u_{-}(t_2+T)\rangle=
 \begin{pmatrix}
 1\\
 0
 \end{pmatrix}
\end{split}.
\end{equation}
We see that indeed $|u_{+}(t_1+T)\rangle=|u_{+}(t_2-T)\rangle$ and $|u_{-}(t_1+T)\rangle=|u_{-}(t_2-T)\rangle$.

In the above adiabatic basis, the transition matrix, referred to as the impulse matrix, can be written as
\begin{equation}
\label{I1}
\begin{split}
I_1=
\begin{pmatrix}
U_{12}& U_{11}\\
U_{22} & U_{21}
\end{pmatrix}
\end{split},
\end{equation}
where the matrix elements are given in Eq.~(\ref{plz4}). Assuming adiabatic evolution between the two LZ transitions, the evolution matrix from $t_1$ to $t_2$ is given by
\begin{equation}
\begin{split}
A_{1\rightarrow 2}=
\begin{pmatrix}
e^{-i{\phi_{d+}(t_2,t_1)}}& 0\\
0& e^{-i{\phi_{d-}(t_2,t_1)}}
\end{pmatrix}
\end{split},
\end{equation}
where ${\phi_{d,\pm}(t_2,t_1)} = \int_{t_1}^{t_2} dt \text{Re}E_{\pm}(t)$ are the dynamical phases accumulated during the adiabatic evolution. Once we determine the impulse matrix $I_2$ at $t_2$ from the Hamiltonian Eq.~(\ref{effham2}), we can obtain the complete evolution matrix as
\begin{align}
\label{UA}
U_A = I_2 A_{1\rightarrow 2} I_1,
\end{align}
which allows us to immediately calculate the final transition probabilities. As we have alluded earlier, the multiple choices for $h_2(t)$ given by Eq.~(\ref{effham2}) suggest that different types of LZS interferometry may exist, depending on the combinations of the coefficients $(\pm m,\pm \gamma)$ in front of the $\sigma_x$ and $\sigma_y$ terms in $h_2(t)$. We now discuss each of these combinations separately in the following.

{\it Case i: $(-m,\gamma)$ and case ii: $(m,-\gamma)$} We consider $(-m,\gamma)$ first. Make use of the transformation $\sigma_y^{-1}h_1(t+t_1)\sigma_y = h_2(t+t_2)$ we obtain
the impulse matrix $I_2$  as
\begin{equation}
\label{I2i}
\begin{split}
I_2=
\begin{pmatrix}
-U_{12}& U_{11}\\
U_{22} & -U_{21}
\end{pmatrix}
\end{split}.
\end{equation}
In the case of $(-m,\gamma)$, the impulse matrix $I_2$ can be directly obtained by letting $-m\rightarrow m$ and $\gamma\rightarrow -\gamma$ in Eq.~(\ref{I2i}). In view of Eq.~(\ref{plz4}), this leads to
\begin{equation}
\label{I2ii}
\begin{split}
I_2=
\begin{pmatrix}
U_{12} & U_{11}\\
U_{22} & U_{21}
\end{pmatrix}
\end{split}.
\end{equation}
Substituting the above $I_2$ matrices and the $I_1$ matrix in Eq.~(\ref{I1}) into Eq.~(\ref{UA}), we find that the transition probabilities for {\it case i} and {\it ii} can be written as
\begin{equation}
\label{c12}
\begin{split}
P_{-+}&=4P_{LZ}(1-P_{LZ})\left (
\sin^2\frac{\phi_t}{2}+\frac{\gamma^2}{m^2-\gamma^2} \right ), \\
P_{+-}&=P_{-+}, \\
P_{--}&=P_{LZ}^2+\left( \frac{m-\gamma}{m+\gamma}\right)^2(1-P_{LZ})^2+\\
&\quad2\frac{m-\gamma}{m+\gamma}P_{LZ}(1-P_{LZ})\cos\phi_t ,\\
P_{++}&=P_{LZ}^2+\left( \frac{m+\gamma}{m-\gamma}\right)^2(1-P_{LZ})^2+\\
&\quad2\frac{m+\gamma}{m-\gamma}P_{LZ}(1-P_{LZ})\cos\phi_t,
\end{split}
\end{equation}
where
\begin{align}
\phi_t = \left \{ \begin{array} {cc} \phi_{d+} - \phi_{d-} + 2\phi_s + \pi & \,{\rm for}\,\, (-m,\gamma) \\
  \phi_{d+} - \phi_{d-} + 2\phi_s &\quad {\rm for}\,\, (m,-\gamma) \end{array} \right. .
\end{align}
{\it Case iii: $(m,\gamma)$ and case iv: $(-m,-\gamma)$} The impulse matrix $I_2$ for $(m,\gamma)$ can again be obtained by letting $-m\rightarrow m$ in Eq.~(\ref{I2i}) and we find
\begin{equation}
\label{I2iii}
\begin{split}
I_2=
\begin{pmatrix}
-U_{21}^*& U_{11}\\
U_{22} & -U_{12}^*
\end{pmatrix}
\end{split}.
\end{equation}
Similarly we obtain the following $I_2$ matrix for $(-m,-\gamma)$ by letting $m\rightarrow -m$ in Eq.~(\ref{I2ii})
\begin{equation}
\label{I2iv}
\begin{split}
I_2=
\begin{pmatrix}
U_{21}^* & U_{11}\\
U_{22} & U_{12}^*
\end{pmatrix}
\end{split}.
\end{equation}
The resulting transition probabilities for both cases can be written as
\begin{equation}
\label{c34}
\begin{split}
P_{-+}&=4\frac{m-\gamma}{m+\gamma}P_{LZ}(1-P_{LZ})
\sin^2\frac{\phi_t}{2}, \\
P_{+-}&=4\frac{m+\gamma}{m-\gamma}P_{LZ}(1-P_{LZ})
\sin^2\frac{\phi_t}{2}, \\
P_{--}&=P_{LZ}^2+(1-P_{LZ})^2+\\
& 2 P_{LZ}(1-P_{LZ})\cos\phi_t, \\
P_{++}&=P_{--},
\end{split}
\end{equation}
where
\begin{align}
\phi_t = \left \{ \begin{array} {cc} \phi_{d+} - \phi_{d-} + 2\phi_s & \,\,{\rm for}\,\, (m,\gamma) \\
  \phi_{d+} - \phi_{d-} + 2\phi_s +\pi &\qquad {\rm for} \,\, (-m,-\gamma) \end{array} \right. .
\end{align}

\begin{figure}
  \centering
  \includegraphics[height=0.32\textwidth]{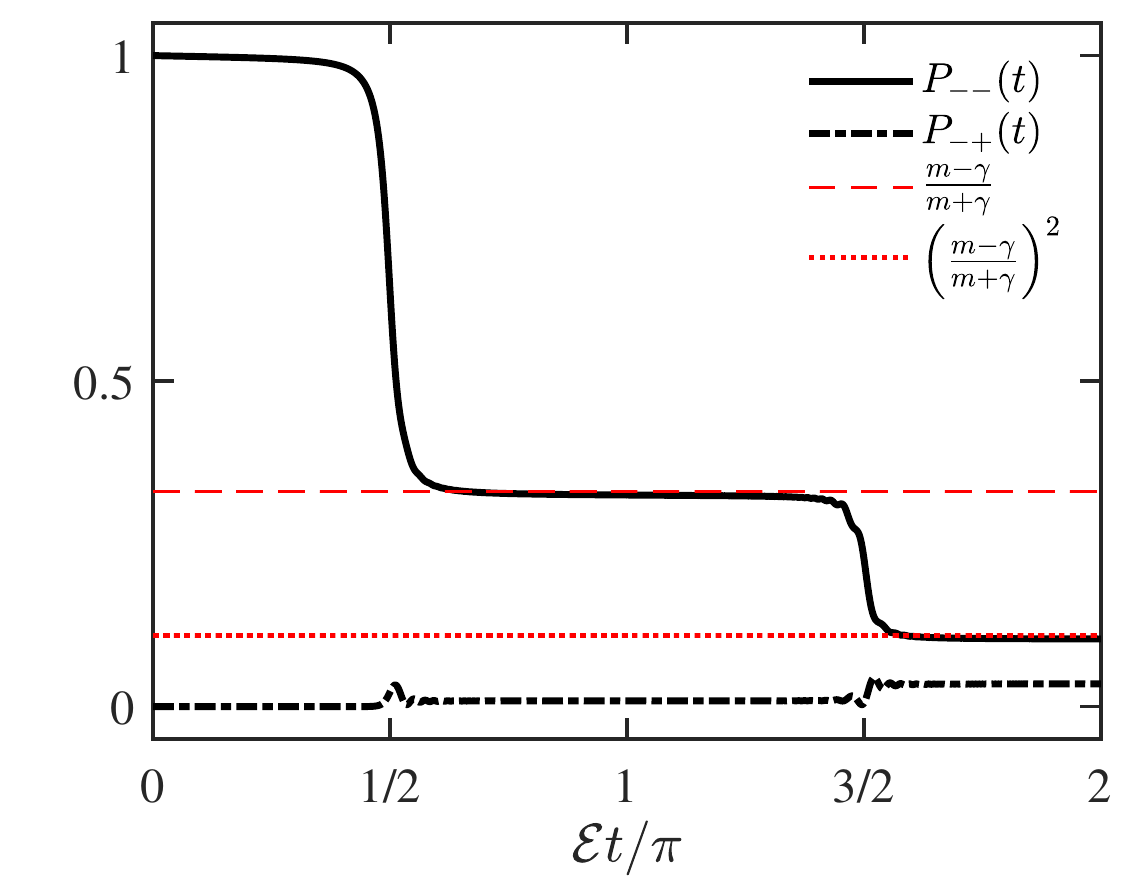}\\
  \caption{Numerically calculated band populations $P_{--}$ and $P_{-+}$ for a non-Hermitian SSH model in the adiabatic limit. The lower band population $P_{--}$ experiences two jumps due to LZ transitions at the first and the second avoided crossing, while the upper band population $P_{-+}$ stays approximately at zero. The parameters are $\alpha=0.1$, $\gamma/J=0.1$ and $\mathcal{E}d/J=0.01$. (see text in Sec.~\ref{sb})}\label{fig6}
\end{figure}

The results in Eqs.~{\ref{c12}} and (\ref{c34}) represent our extension of the Hermitian LZS transition probability given in Eq.~(\ref{LZS0}) to non-Hermitian systems with real adiabatic parameters. The presence of the non-Hermitian term gives rise to several properties unique to the non-Hermitan system. First, the number of distinctive LZS interferometry doubles compared to the corresponding Hermitian system, all of which recover Eq.~(\ref{LZS0}) by letting $\gamma\rightarrow 0$. Second, the transition probabilities are again not fully symmetrical with respect to the different initial states, which is to be expected from earlier results of the non-Hermitian LZ transition. For {\it case i} and {\it case ii}, the inter-band transition probabilities are symmetrical while the intra-band ones are not; the situation is reversed for {\it case iii} and {\it case iv}. In fact, further contrast between {\it case i,ii} and {\it case iii,iv} exist. For the former, the inter-band transition probabilities are finite for any non-zero sweep velocity, while those for the latter vanish at $\phi_t = 2j\pi$ $(j = 0,1,2,\cdots)$, which is analogous to Hermitian system. The similarity between the non-Hermitian {\it case iii,iv} and the Hermitian system is also reflected in the adiabatic limit $\delta\rightarrow \infty$, where the driven particles stay in the same adiabatic band with full population during the evolution. In contrast, the band population for {\it case i,ii}  jumps twice at two avoid level crossings and reaches the value $(m-\gamma)^2/(m+\gamma)^2$ for the spin-down state and  $(m+\gamma)^2/(m-\gamma)^2$ for the spin-up. This is illustrated in Fig. \ref{fig6}, where we numerically calculate the transition probabilities for a non-Hermitian SSH to be discussed in Sec.~\ref{sb}. Such a property was first noticed in Ref. \cite{ptlz} from a numerical calculation.

Despite significant differences between the non-Hermitian and Hermitian LZS interferometry, an important common feature remains. In the Hermitian system, two types of LZS interferometry exist, distinguished by the relative sign of the mass terms at the two level crossings.  The difference in the corresponiding LZS transition probabilities  is characterised by a $\pi$ difference in the total accumulated phase $\phi_t$, which can be seen in Eqs.~{\eqref{c12}}  with $\gamma =0$. As discussed in Ref. \cite{lzspiphase,lzs2,lzs4}, such a $\pi$ phase can be derived through the concept of open path geometric phase. Here we find that the $\pi$ phase is retained in the presence of non-Hermiticity. As we can see from Eqs.~(\ref{c12}) and (\ref{c34}), the appearance of the $\pi $ phase only depends on the relative sign of the mass terms, meaning that it survives the change of non-Hermitian parameter from $\gamma$ to $-\gamma$. Such a robustness is consistent with its geometric origin.

\subsection{Application to the non-Hermitian SSH model}
\label{sb}
The non-Hermitian SSH model has received much interest recently due to the important role it plays in the development of the non-Hermitian topological band theory. Here we focus on the dynamical aspects of the model and study the Bloch oscillation of a particle driven by a constant electric field~\cite{nonBO1,nonBO2,nonlzR}. Since the energy spectrum of the non-Hermitian SSH model features two identical level crossings with opposite gap parameters in the reciprocal space, the dynamics of the driven particle realises precisely the previously discussed {\it case i} non-Hermitian LZS interferometry.  We consider the following non-Hermitian SSH model under a uniform electric field $\mathcal{E}$
\begin{align}
\label{nSSH}
H&=\sum_{i=1}^N(J_1c_{1,i}^\dagger c_{2,i}+J_2c_{2,i}^\dagger c_{1,i+1}+\text{h.c.}) \nonumber \\
&+\sum_{i=1}^N i\gamma (c^\dagger_{1,i}c_{1,i}- c^\dagger_{2,i}c_{2,i}) \nonumber \\
&-\mathcal{E}\sum_{i=1}^N(x_{1i}c_{1,i}^\dagger c_{1,i}+x_{2,i}c_{2,i}^\dagger c_{2,i}),
\end{align}
where $J_{1}$ and $J_2$ respectively denote the hopping parameters between the intra-cell and inter-cell nearest neighbour sites, and $x_{1,i}$ and $x_{2,i}$ are lattice vectors of the two sites in the $i$-th cell. Here the second line is a non-hermitian term describing gain on one sublattice and  loss on the other.  For this reason we refer to Eq.~(\ref{nSSH}) as the gain-and-loss SSH model. Here we assume that the lattice sites are uniformly spaced at a distance $d$ and the nearest neighbour sites within one sublattice is $2d$. By introducing a gauge transformation
 \begin{figure}
  \centering
  \includegraphics[height=0.31\textwidth]{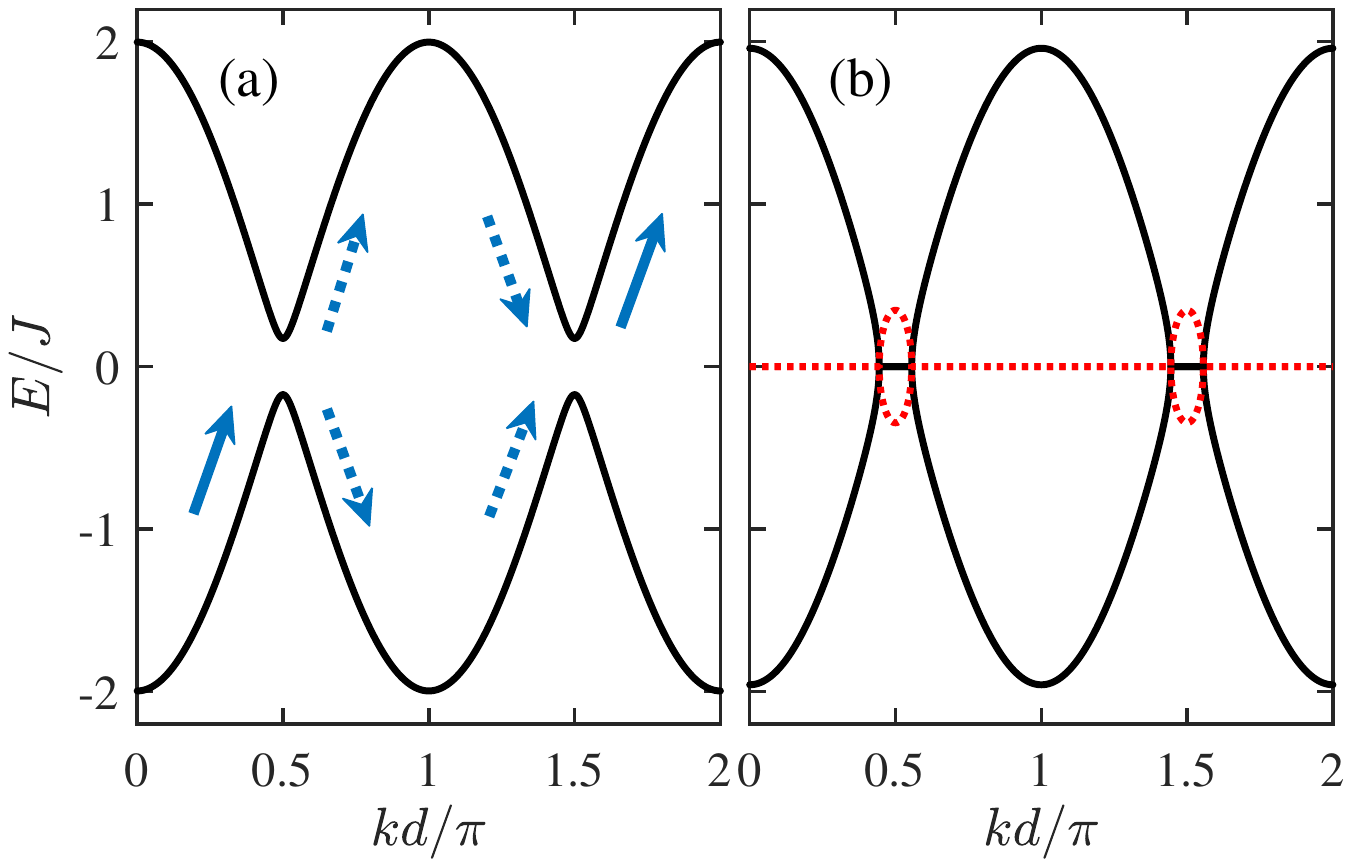}\\
  \caption{The adiabatic energy spectrum of the non-Hermitian SSH model with its Hamiltonian given by Eq.~\eqref{ham2}. (a) Energy spectrum for $|\gamma|<|2\alpha J|$ and an illustration of the LZS interferometry. The initial state is prepared at the lower band. Driven by the external force, the state splits up into two at the first avoided crossing and then interfere at the second. The adiabatic-impulse theory assumes that the whole evolution can be decomposed as an impulse transition at the anti-crossing points and adiabatic evolution everywhere else. (b) For $|\gamma|>|2\alpha J|$, exceptional points emerge and part of the energy spectrum becomes complex. The solid and dashed lines denote the real and imaginary part of the energy spectrum, respectively.  }\label{lzs}
\end{figure}
\begin{equation}
V(t)=e^{ -i\mathcal{E}t\sum_{i=1}^N(x_{1,i}c_{1,i}^\dagger c_{1,i}+x_{2,i}c_{2,i}^\dagger c_{2,i})},
\end{equation}
we obtain an equivalent Hamiltonian
\begin{equation}
\begin{split}
H_V&=\sum_{i=1}^N(J_1e^{ i\mathcal{E}t} c_{1,i}^\dagger c_{2,i}+ J_2 e^{ i\mathcal{E}t}c_{1,i+1}c_{2,i}^\dagger+\text{h.c.})\\
&+\sum_{i=1}^N\left[i\gamma (c^\dagger_{1,i}c_{1,i}- c^\dagger_{2,i}c_{2,i})\right],
\end{split}
\end{equation}
which recovers the discrete translational symmetry. In Bloch basis, the above Hamiltonian can be written as $H_V=\sum_k\psi^\dagger_kh (k)\psi_k$, where $\psi_k =(c_{1k},c_{2k})^T$ and the Bloch Hamiltonian $h(k)$ reads
\begin{equation}
\label{ham2}
h(k)=2J\cos (kd+\mathcal{E}t)\sigma_x-2\alpha J\sin (kd+\mathcal{E}t)\sigma_y+i\gamma\sigma_z.
\end{equation}
Here $J_{1,2}=J(1\pm \alpha)$ and $k$ denotes the quasimomentum in the first Brouillon zone $[0, \pi/d]$. In absence of the electric field, the energy band spectrum is
\begin{align}
E_{\pm}(k)=\pm\sqrt{4J^2\cos^2(kd)+4\alpha^2J^2\sin^2(kd)-\gamma^2 },
\end{align}
 which is real for $|\gamma|<|2\alpha J| $ and features two avoided level crossings at $kd = \pi/2,3\pi/2$, as shown in Fig.~\ref{lzs} (a).  When $|\gamma|=|2\alpha J|$, the bands touch at $kd=\pi/2$ which gives rise to the so-called exceptional point. Two pairs of exceptional points appear in the energy spectrum for $|\gamma|>|2\alpha J|$ and the energy is purely imaginary between each pair of them. Such a regime is referred to as the $\mathcal{PT}$-symmetry-broken regime and a typical spectrum is shown in Fig.~\ref{lzs} (b).

\begin{figure}
  \centering
  \includegraphics[height=0.35\textwidth]{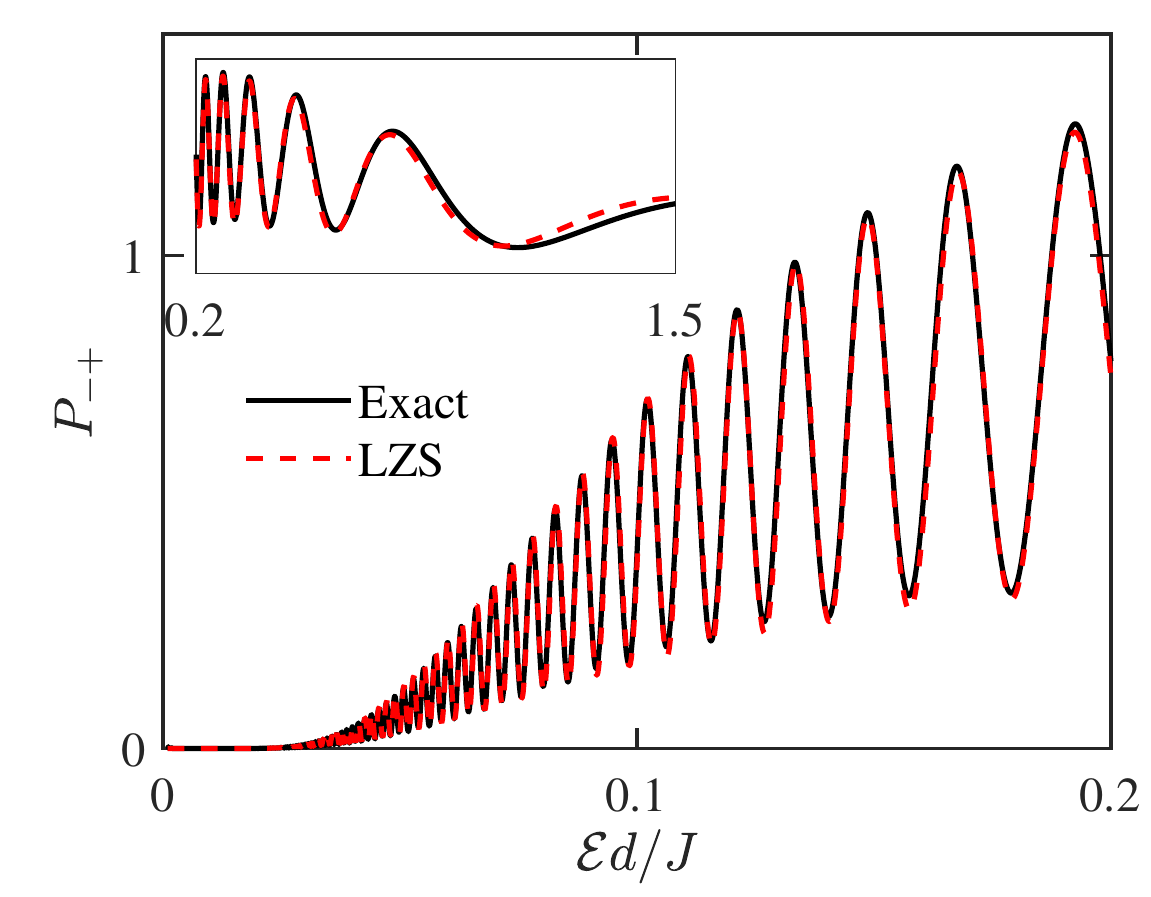}\\
  \caption{(Color Online) Transition probability $P_{-+}$  as a function of $ \mathcal{E}$ for the non-Hermitian SSH model in  the $\mathcal{PT}$ symmetry preserving regime, i.e., $\delta>0$. The solid line is from numerically solving the Schr\"{o}dinger's equation  and the dashed line is from the adiabatic-impulse theory of the LZS interferometry.  The results from the two calculations agree with each other very well. Slight deviations appear for large $\mathcal{E}$ (see inset). The parameters are $J=1$, $\alpha=0.2$ and $\gamma/J=0.2$. }\label{fig7}
\end{figure}

 \begin{figure}
  \centering
   \includegraphics[height=0.35\textwidth]{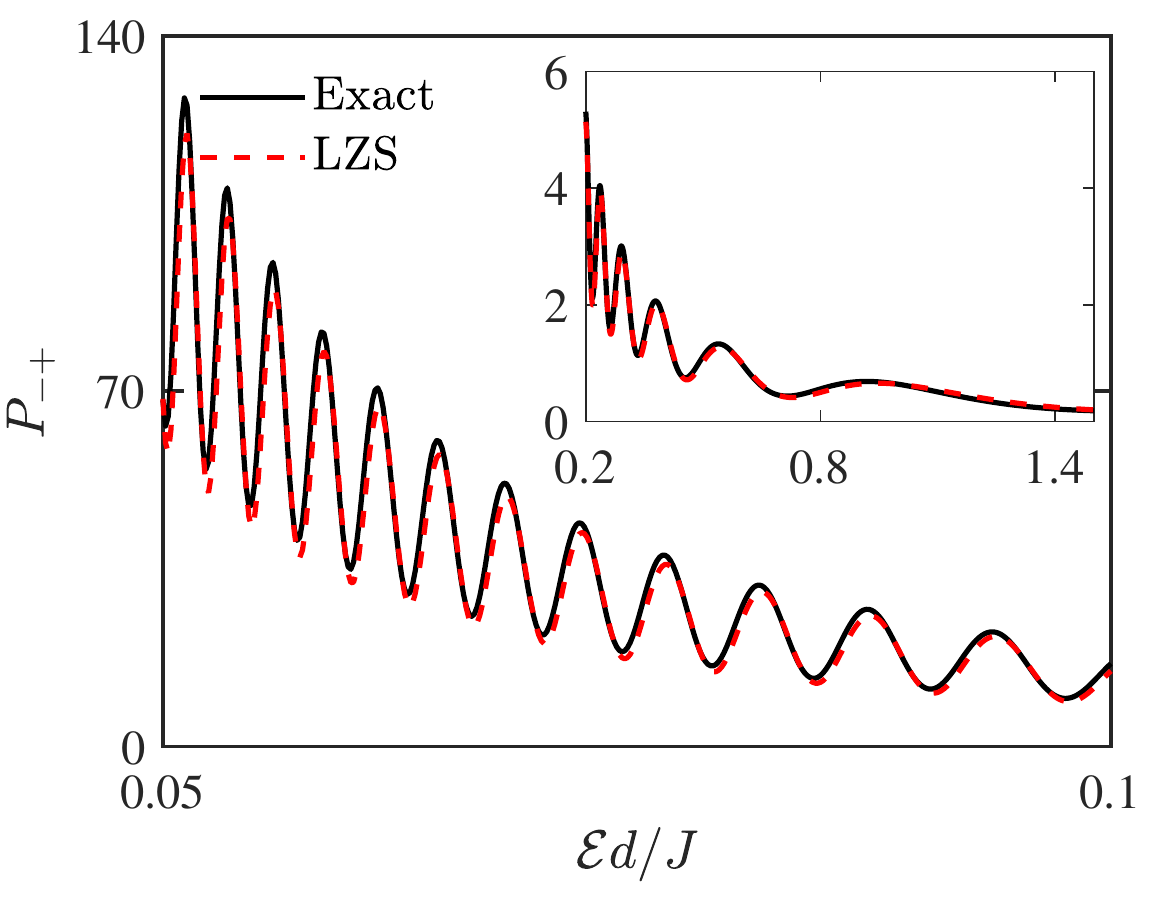}\\
  \caption{(Color Online) Transition probability $P_{-+}$  as a function of $ \mathcal{E}$ for the non-Hermitian SSH model in  the $\mathcal{PT}$ symmetry broken regime, i.e., $\delta<0$. The deviation of  LZS result from the exact one increases in the ${\mathcal E} \rightarrow 0$ limit, i.e., $\delta\rightarrow -\infty$. As explained in Eq. \eqref{taulz}, this is due to the fact the LZ transition time increases exponentially in such a limit. The parameters in the numerical computation are $J=1$, $\alpha=0.1$ and $\gamma/J=0.3$. }\label{figs8}
\end{figure}
 Consider a particle initially located at $k = 0$ in the lower band. Under the electric field, its wave function evolves according to Eq.~(\ref{ham2}) and undergoes the Bloch oscillation. To see the connection with the LZS interferometry discussed earlier, we perform a rotation $(\sigma_x,\sigma_y,\sigma_z)\rightarrow(-\sigma_z,-\sigma_x,\sigma_y$), expand the Hamiltonian at $ t_1= \pi/2{\mathcal E}$ and $t_2= 3\pi/2{\mathcal E}$, and arrive at
\begin{align}
h_1(t)&= 2J{\mathcal E}(t-t_1)\sigma_z+2\alpha J\sigma_x+i\gamma \sigma_y \\
h_2(t)&= -2J{\mathcal E}(t-t_2)\sigma_z-2\alpha J\sigma_x+i\gamma \sigma_y.
\end{align}
Identifying $F = 2J{\mathcal E}$ and $m =  2\alpha J$, we can immediately see that the quantum dynamics realises the previously discussed {\it case i} non-Hermitian interferometry.

In Figs.~\ref{fig7} and \ref{figs8}, we plot the LZS transition probability $P_{-+}$ calculated from Eq.~(\ref{c12}) as a function of the electric filed $\mathcal{E}$ for positive and negative adiabatic parameters respectively. We see that the comparisons of our analytic results with the exact dynamics based on Eq.~(\ref{ham2}) are good. The slight deviation of our LZS results in the case of negative $\delta$ can be explained by the fact that the LZ transition time $\tau_{LZ}$ increases exponentially as $\delta\rightarrow -\infty$. As a result, it takes longer for the LZ transition probabilities to reach their asymptotic values after the level crossing, which renders the adiabatic-impulse model less accurate.

\section{Experimental simulation}
\label{s4}
In this section we propose an experimental scheme using photonic waveguide arrays to simulate the dynamics of the non-Hermitian SSH model discussed in Sec.~\ref{sb}.
In photonic waveguide arrays, the $\mathcal{PT}$-symmetric non-Hermitian lattice model can be realised with a complex potential \cite{pt1,pt2} or by alternatively bending the waveguides along the propagation direction \cite{gainandloss,Weimann2016}. Similarly the external linear potential can be imposed either by enforcing a refractive index gradient in the transverse direction \cite{traverse1,traverse2} or by adding a geometric curvature of the waveguides \cite{bend1,bend2} in the propagation direction. Using these methods, the non-Hermitian SSH model has been realised experimentally in a passive way, meaning that only a loss effect is exerted on either set of sublattice. However, a simple transformation allows one to deduce the dynamics of the gain-and-loss model from the passive one. In order to make connections with our results in earlier sections,  we focus on the the gain-and-loss photonic waveguide arrays. 

\begin{figure}
  \centering
  \includegraphics[height=0.32\textwidth]{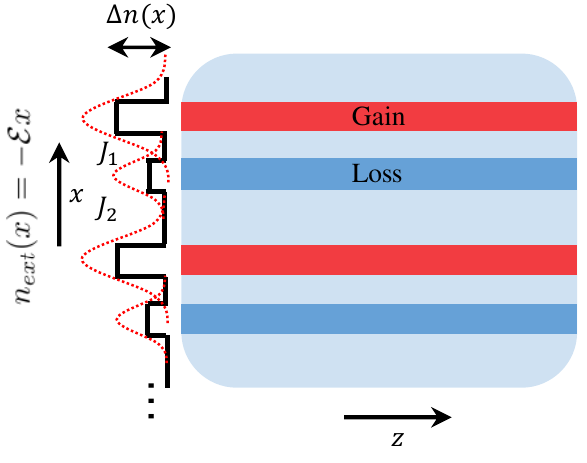}\\
  \caption{Sketch of photonic waveguide arrays for the experimental simulation of non-Hermitian LZS interferometry. }\label{setup}
\end{figure}

For concreteness, we take the propagation direction of the waveguides to be along the $z$ direction and the array is arranged along the $x$-direction (see Fig.~\ref{setup} for a sketch of the experimental system). For a light beam with wavelength $\lambda$ trapped in the waveguides, the electrical field envelope $E(x,z)$ is described by the optical paraxial Helmholtz equation~\cite{cm1}
\begin{equation}
\label{Helmholtz}
i\tilde \lambda \frac{\partial}{\partial z}E(x,z)=-\left(\frac{\tilde \lambda ^2}{n_0}\frac{\partial^2}{\partial x^2}+\Delta n(x) + n_{ext}(x)\right)E(x,z),
\end{equation}
where $\tilde \lambda = \lambda/2\pi$ is the reduced wavelength, $\Delta n(x)\equiv n_0-n(x)$ is the periodical variation of the refractive index $n(x)$ relative to the bulk value $n_0$ and  $n_{ext}(x) = - \mathcal{E} x$ is the refractive index gradient imposed, for instance, through a temperature gradient in the traverse direction.  The above equation shares a similar form with the time-dependent Schr\"{o}dinger's equation of a one-dimensional lattice system and the length of the waveguide plays the role of the time scale. Due to this similarity, we expand $E(x,z)$ in terms of the confined modes trapped in waveguides   $w(x-x_m)$ 
\begin{align}
E(x,z) = \sum_m \psi_m(z) w(x-x_m),
\end{align}
where $\psi_m(z)$ is the modal amplitude and $x_m$ is location of the $m$-th waveguide along the $x$ direction. Applying the tight-binding approximation and considering the additional gain and loss described by the parameter $\gamma$, we find that the modal distribuation of the photonic field satisfies the following equation
\begin{equation}
\label{rspaceEq}
i\frac{d \psi_m}{dz}+J_m e^{i\varphi}\psi_{m+1}+J_m e^{-i\varphi}\psi_{m-1}-i\gamma(-1)^{m} \psi_m=0,
\end{equation}
where $\varphi(z) = -\mathcal{E}z $ and $J_m$, obtainable from Eq.~(\ref{Helmholtz}), denote the coupling rate between adjacent waveguides.  Engineering the refractive index $n(x)$ such that $J_m$ alternate between $J_1$ and $J_2$ on adjacent lattice sites, the above equation is then mathematically equivalent to the time-dependent Schr\"{o}dinger equation based on the Hamiltonian Eq.~(\ref{nSSH}). 

\begin{figure}
  \centering
  \includegraphics[height=0.32\textwidth]{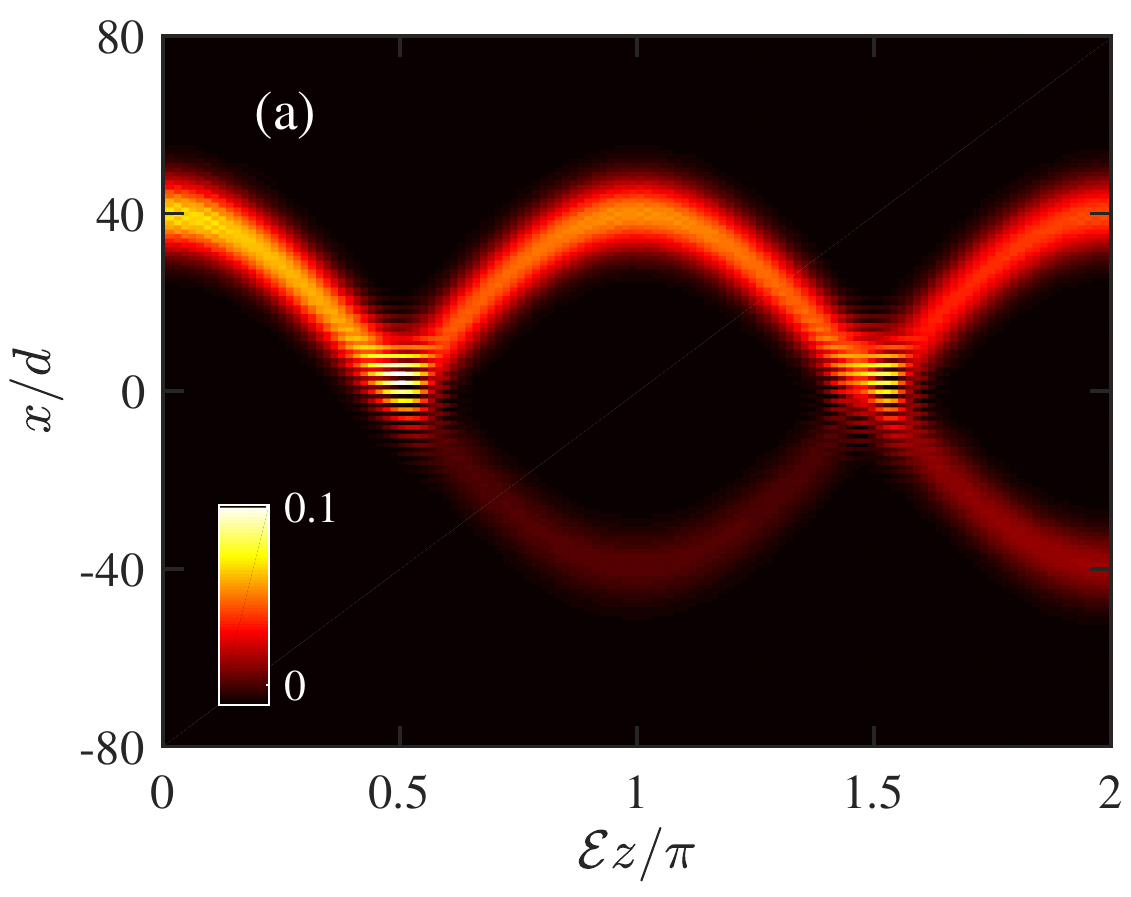}\\
  \includegraphics[height=0.32\textwidth]{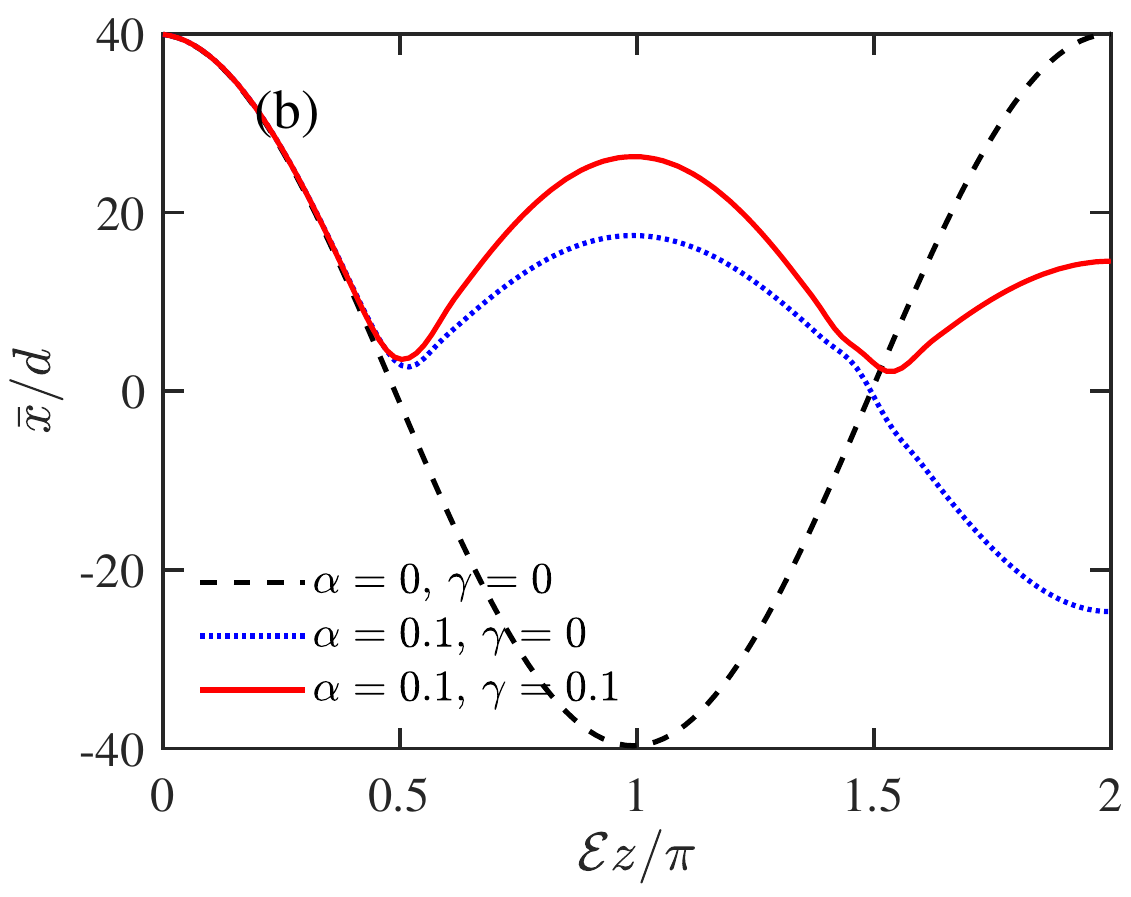}
  \caption{(Color Online)  (a)  Spatial configuration of the renormalised beam profile calculated from Eq.~\eqref{rspaceEq}. The profile moves along the negative $x$-direction at first and splits up into two at $\mathcal{E}z=\frac{1}{2}\pi$. Here the parameters are  $\alpha=0.1$ and $\gamma/J=0.1$. (b) Numerically calculated CoM of beam density profile versus ``time'' $z$, where the beam is initially centred at the $40$-th site. In the simulation the number of the waveguides is $160$ and the width of the initial wave packet is $8d$. The force or electric field is $\mathcal{E}=0.05$.  }\label{com}
\end{figure}

As a result of the simulated electric field, the central position of the modal amplitude profile changes along the direction of the waveguide, which precisely simulates the Bloch oscillation of the electron in the SSH model. In Fig.~\ref{com} (a) we plot the intensity profile as a function of the $z$ coordinate, where the modal amplitude profile at the boundary of the array is chosen to be a Gaussian beam
\begin{equation}
\psi_m(z=0)=\frac{1}{\sqrt{\sqrt{\pi}l}}e^{-\frac{(x_m-x_0)^2}{2l^2}}.
\end{equation}
Here $l$ is the width of the profile and $x_0$ is the position of its centre. In the broad limit where the width of the profile is much larger than the lattice spacing, the beam profile at $z=0$ corresponds to a supposition of upper band Bloch states in the close vicinity of $k =0$~\cite{nonlzR}. Thus, under the application of the ``electric field'', the beam profile in the Fourier space moves along the upper band and undergoes two LZ transitions at $z = \pi/2\mathcal{E}$ and $z = 3\pi/2\mathcal{E}$. Such transitions can be vividly seen in real space (see Fig.~\ref{com} (a)), where the initial single wave packet splits up into two at $z = \pi/2\mathcal{E}$, which meet again at $z = 3\pi/2\mathcal{E}$.

 To experimentally test our analytic results for the non-Hermitian LZ transition and LZS interferometry, one can vary the ``electric field'' $\mathcal {E}$ and measure the centre of mass (CoM) of the total intensity profile at $z= \pi/\mathcal{E}$ and $z= 2\pi/\mathcal{E}$ defined as
 \begin{align}
 \bar x(z) =\frac{ \sum_m x_m |\psi_m(z)|^2}{ \sum_m |\psi_m(z)|^2}.
 \end{align}
 By projecting the profile into the upper and lower band components, the CoM of each component can be determined from the Ehrenfest theorem. At $z=\pi/\mathcal{E}$ the displacement of the upper band component is zero while that of the lower band is $\Delta x=-E_w/\mathcal{E}$, where $E_w=2E_+(k=0)$ denotes the band width. Thus the CoM of the total profile at $z=\pi/\mathcal{E}$ is the sum of the two components weighted by the transition probabilities
\begin{equation}
\bar{x}(z=\pi/\mathcal{E})=\frac{(x_0+\Delta x) |U_{22}|^2+x_0|U_{12}|^2}{|U_{22}|^2+|U_{12}|^2},
\end{equation}
where the matrix elements are given in Eq.~(\ref{plz4}) with the substitution $F = 2J\mathcal{E}$ and $m = 2\alpha J$. Similarly, we can determine the CoM of the profile at $z=
2\pi/\mathcal{E}$ as
\begin{equation}
\label{com2pi}
\bar{x}(z=2\pi/\mathcal{E})=\frac{(x_0+\Delta x)P_{+-}+x_0P_{++}}{P_{+-}+P_{++}},
\end{equation}
where the transition probabilities are given in Eq.~(\ref{c12}). An example of the CoM trajectory $\bar x (z)$ is shown in Fig.~\ref{com} (b).

\begin{figure}
  \centering
  \includegraphics[height=0.32\textwidth]{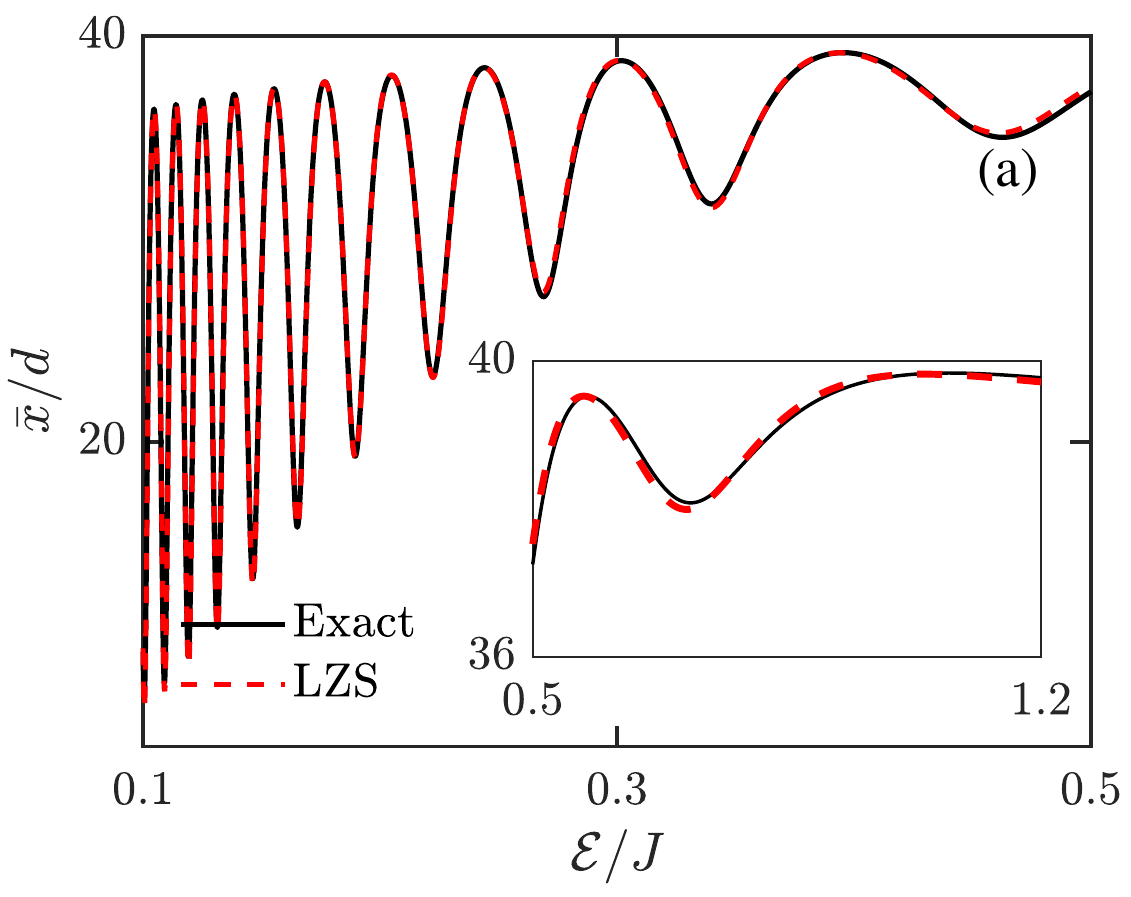}\\
  \includegraphics[height=0.32\textwidth]{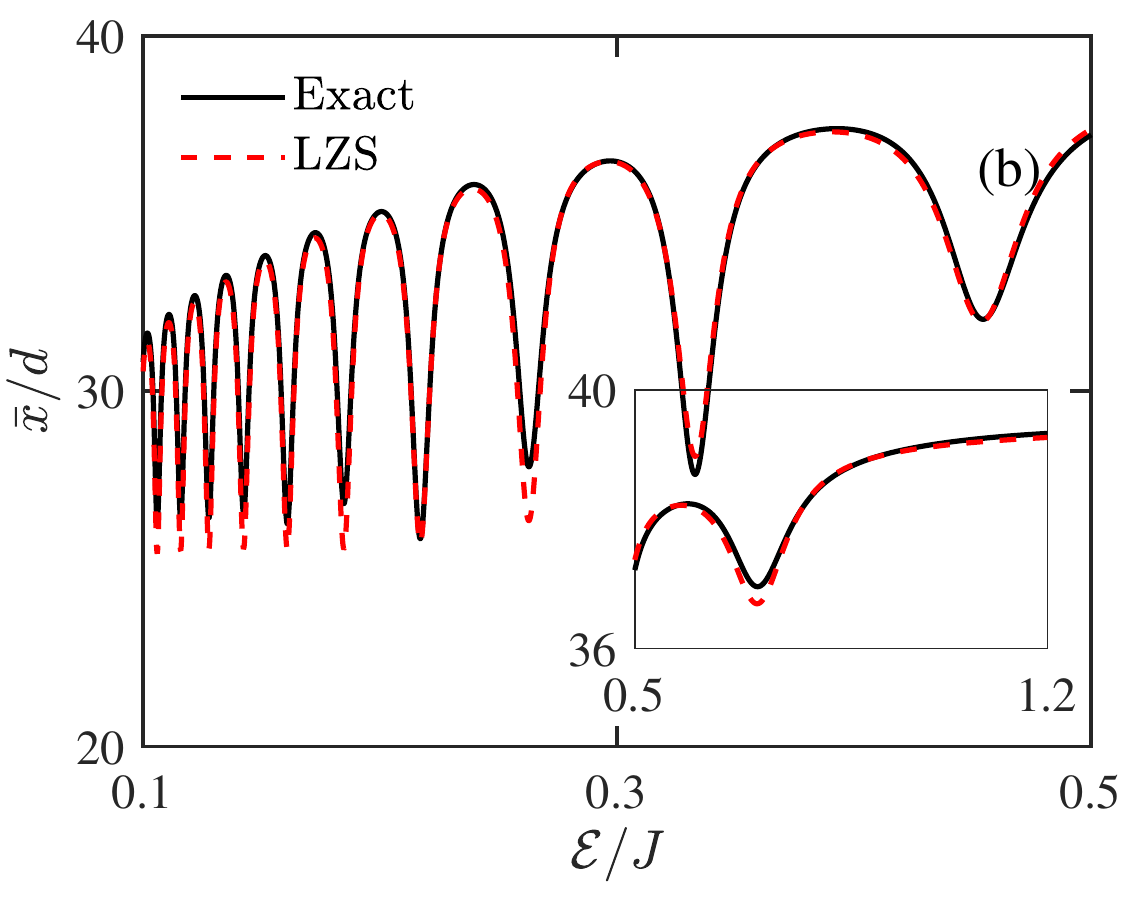}\\
  \caption{The CoM of beam at time $z=2\pi/\mathcal{E}$ versus the force $\mathcal{E}$. The solid line is from solving Eq.~(\ref{rspaceEq}) numerically and the dashed line is from Eq.~(\ref{com2pi}). (a): Excellent agreement is exhibited when the adiabatic parameter $\delta>0$. (b): The LZS result deviates slightly from the exact one for $\delta<0$. In the simulation, the total number of sites is 240 and the initial profile is centred at the $40$-th site. The rest of the parameters are (a) $\alpha=0.1$, $\gamma/J=0.1$ and (b) $\alpha=0.1$, $\gamma/J=0.3$. }\label{com_LZS}
\end{figure}

In Fig. \ref{com_LZS}, we plot the CoM of the profile as a function of $\mathcal{E}$ at  $z=
2\pi/\mathcal{E}$. Comparison with the exact results obtained by numerically solving Eq.~\eqref{rspaceEq} are also shown. For the parameters chosen, we see that the analytic LZS results are in excellent agreement with the numerics, especially for a positive adiabatic parameter $\delta$.  Even for negative adiabatic parameters, which indicate the ${\mathcal P}{\mathcal T}$-symmetry-breaking regime, the adiabatic-impulse model description is still reasonably good. However, in this regime the LZ transition time increases exponentially as $\delta$ decreases. Thus we would expect that the deviation of the LZS result from the exact dynamics would deteriorate as $\delta =[ {(2\alpha J)^2 - \gamma^2}]/{2J{\mathcal E}} \rightarrow -\infty$, which is precisely what we can observe from Fig. \ref{com_LZS} (b).

\section{Conclusion}
\label{s5}
In summary, we have obtained solutions to the LZ transition for a general non-Hermitian Hamiltonian and have analysed the LZS interferometry for $\mathcal{PT}$ symmetric non-Hermitian systems. In the non-Hermitian LZ solutions, the adiabatic parameter can be complex and the presence of the imaginary part may drastically modify the transition probabilities. Focusing on systems with a real adiabatic parameter, we have formulated the non-Hermitian LZS interferometry using the adiabatic-impulse theory and derived analytic formulae for the final transition probabilities. On the one hand, significant differences arise between the non-Hermitian LZS interferometry and its Hermitian counterpart. One of them is that the number of different types of interferometry increases due to the non-Hermiticity. Another is that the adiabatic parameter can be negative for the non-Hermitian system, which has no equivalence in the Hermitian case. On the other hand, certain features of the Hermitian LZS transition, the geometrical $\pi$ phase shift in particular, persist in the non-Hermitian system. Finally, we show that the Bloch oscillation in a $\mathcal{PT}$ symmetric non-Hermitian SSH model naturally realises one type of the non-Hermitian LZS interferometry we analysed. Our proposal of using photonic waveguides array to simulate this kind of dynamics may be of interest to experimentalists in this field.

\section*{Acknowledgements}
X. S. and Z. W. acknowledge supports by NSFC (Grants No. 11974161) and Key-Area Research and Development Program of GuangDong Province (Grant No. 2019B030330001). F. W. acknowledges support by NSFC (Grants No. 11904159). Z. L. acknowledges support by  NSFC (Grants No. 11704132).

\emph{Note added.}---Some of the problems addressed in this paper have also been studied in a recent work by Longstaff and Graefe~\cite{nonlzR}, and our results agree wherever the overlap occurs..

\bibliographystyle{apsrev4-1}
\bibliography{ref_nssh}
\end{document}